# Effect of damage by 2-MeV He ions on the normal and superconducting properties of magnesium diboride

R. Gandikota,[a)] R.K. Singh,[a)] J. Kim,[a)] B. Wilkens,[b)] N. Newman,[a),b),f)] J.M. Rowell[a)], A.V. Pogrebnyakov,[c),d),e)] X.X. Xi,[c),d),e)] J.M. Redwing,[d),e)] S.Y. Xu,[c),e)] and Qi Li[c),e)]

[a)] Department of Chemical and Materials Engineering, Arizona State University, Tempe, Arizona 85287-6006

[b)] Center for Solid State Science, Arizona State University, Tempe, Arizona 85287-1704

[c)] Department of Physics, The Pennsylvania State University, University Park, Pennsylvania 16802.

[d)] Department of Materials Science and Engineering, The Pennsylvania State University, University Park, Pennsylvania 16802.

[e)] Materials Research Institute, The Pennsylvania State University, University Park, Pennsylvania 16802.

[f)] Author to whom correspondence should be addressed;
   Electronic mail: Nathan.Newman@asu.edu

**Abstract**

We have studied the effect of damage induced by 2-MeV alpha particles on the critical temperature, $T_c$, and resistivity, $\rho$, of $MgB_2$ thin films. This technique allows defects to be controllably introduced into $MgB_2$ in small successive steps. $T_c$ decreases linearly as the intragrain resistivity at 40 K increases, while the intergrain connectivity is not changed. $T_c$ is ultimately reduced to less than 7 K and we see no evidence for a saturation of $T_c$ at about 20 K, contrary to the predictions of the $T_c$ of $MgB_2$ in the dirty limit of interband scattering.

MgB$_2$ is the first superconductor to exhibit two distinct and widely different energy gaps, arising from a three dimensional π-band, and a two-dimensional σ-band of carriers, and its resistivity (ρ), transition temperature (T$_c$), and critical field (H$_c$) are of particular interest. The normal-state resistivity of MgB$_2$ samples from various groups is found to vary over a very wide range of values. The resistivity of samples with T$_c$s close to 39 K varies from < 1 μΩ.cm to > 100 mΩ.cm.[1,2] While this has been cited as an indication of weak interband scattering,[3] a simple model[4] ascribed such large resistivity changes primarily to decreased connectivity between the grains, created for example by intergrain impurities such as MgO, BO$_x$ or boron carbides. The grains themselves can be relatively clean even when they are poorly connected. The concurrent reduction in critical currents in high resistivity MgB$_2$ films supports this simple model.[5] No correlation exists between T$_c$ and H$_{c2}$ with the resistivity of as-made MgB$_2$ samples.[6] It has been predicted that in the dirty limit of interband scattering, T$_c$ should saturate around 20 K.[7,8] In this letter, we show that introducing disorder in small sequential steps into low-resistivity films offers distinct advantages in the study of scattering processes in the multiband superconductor MgB$_2$. Understanding of such scattering processes has taken on new importance with the observation of very high H$_{c2}$ values in the C-doped MgB$_2$ films.[8] We use α-particle irradiation, which generates predominantly point defects, to controllably alter the properties of MgB$_2$ films so that we can differentiate the mechanism(s) responsible for the wide range of resistivities found in MgB$_2$ samples. We conclude that damage increases the residual resistivity of the grains themselves but has little or no effect on the connectivity of initially clean films, and that T$_c$ is reduced to ~7 K as the residual resistivity increases, but no saturation of T$_c$ due to interband scattering in dirty MgB$_2$ is observed.

Two MgB$_2$ thin films (A and B) were grown at PSU on 5 mm × 5 mm sapphire substrates using hybrid physical-chemical vapor deposition.[9] To facilitate lithography on the sample and to minimize moisture affecting the MgB$_2$ film, a Ta$_x$N film (~ 400 Å thick) was deposited by sputtering on top of the superconducting film. Such Ta$_x$N films have a resistivity of 1.9 mΩ.cm at room temperature and 11 mΩ.cm at 4.2 K; hence the Ta$_x$N layer has a resistivity high enough to not significantly influence the electrical measurements of the MgB$_2$-Ta$_x$N bilayer films while still allowing contacts to be made easily. The MgB$_2$ films in samples A and B were 650 Å and 2500 Å thick, respectively, determined using Rutherford Backscattering Spectrometry (RBS). The bilayer film was patterned into a 100 μm × 1 mm bridge using reactive ion etching to define the area of damage and to provide for four-point measurements. The values of Δ$\rho_{300\text{-}41K}$ [=$\rho$(300K)-$\rho$(41K)] for films A and B after patterning were 11 and 7.4 μΩ.cm. The Δ$\rho_{300\text{-}41K}$ value of film A is somewhat higher than the typical values for PSU films[9] and bulk samples[1] (between 7 and 9 μΩ.cm), which may be attributed to non-optimal growth conditions, degradation, Ta$_x$N passivation, and/or the bridge patterning.

The α-particle irradiation was carried out at room temperature using a 2-MeV $^4$He$^{++}$ ion beam from a Tandem accelerator. The MgB$_2$ and the Ta$_x$N layers were thin enough so that the 2-MeV $^4$He$^{++}$ ions damage the entire film cross-section uniformly. The sample was tilted 8º to avoid possible ion channeling. An α-particle beam of 3.5 mm × 3.5 mm size was used to ensure uniform irradiation over the entire 1 mm length of the bridge. A total dose of 1.3×10$^{17}$ cm$^{-2}$ was achieved over 11 steps of irradiation, taking a total time of ~ 35 hours. The resistivity was measured before irradiation and after each irradiation step using a dipping probe and a Quantum Design Physical Property Measurement System (PPMS).

For the film A, Fig.1 shows the temperature dependence of resistivity ($\rho$) before irradiation and after selected irradiation steps. T$_c$ decreases monotonically while the resistivities $\rho$(300K) and $\rho$(41K) increase markedly with damage but Δ$\rho_{300\text{-}41K}$ decreases by only 30 %. In the review of Reference 4, Δ$\rho_{300\text{-}41K}$ was defined as the change in resistivity from 300 K to a temperature just above $T_c$, 41 K in this case. It was also pointed out that an increase in $\Delta\rho_{300-41K}$ is most readily explained by a decrease in the connectivity of the sample (the more commonly used residual resistivity ratio is not a measure of connectivity). An increase in the residual resistivity $\rho(41K)$ can be due to an increase in the intragrain scattering and/or to a reduction in connectivity, while an increase in $\Delta\rho_{300-41K}$ is a measure of a decrease in intergrain connectivity (e.g., a decrease in the sample cross-sectional area that carries the current, due to porosity or intergrain effects). Samples with highest $T_c$ and lowest residual resistivity are assumed to be fully connected and their $\Delta\rho_{300-41K}$s are usually between 7 and 9 $\mu\Omega$.cm.[1,9]

Our observation that $\rho(41K)$ increases from 1.9 $\mu\Omega$.cm to 100 $\mu\Omega$.cm, while $\Delta\rho_{300-41K}$ remains relatively constant, implies that the ion beam damage increases only the intra-grain scattering, presumably as a result of the generation of point defects that reduce the electron mean free path. The connectivity of the sample, as inferred from $\Delta\rho_{300-41K}$, is not affected by the damage. The small decrease in $\Delta\rho_{300-41K}$ can be explained by the "saturation resistivity" model used to describe A15 superconductors.[10,11]

Our observations differ from the results of irradiation of $MgB_2$ films by $10^{12}$ cm$^{-2}$ 200-MeV Ag ions, in which $\Delta\rho_{300-41K}$ nearly doubled after irradiation while there was little change in $T_c$.[12] The dissimilarities between the two results may be attributed to the large difference in the average energy imparted to the target atoms. A dose of $10^{17}$ cm$^{-2}$ 2-MeV He$^{++}$ ions, as used in our study, results in an imparted energy per unit volume[13,14] that is about 2000 times greater than that with $10^{12}$ cm$^{-2}$ 200-MeV Ag ions.[8]

Figure 2 shows the variation of $T_{c\ onset}$ ($T_{co}$), $T_{c\ finish}$ ($T_{cf}$), $\rho(41K)$, and $\Delta\rho_{300-41K}$ with the $^4$He$^{++}$ ion dosage. $T_{co}$ is defined as the temperature where there is a noticeable increase in the slope of $\rho$ vs. T plot and $T_{cf}$ is defined as the temperature where the resistance goes below the noise level of the multimeter, as temperature is reduced. Both $T_{co}$ and $T_{cf}$ reduce monotonically with dosage, with the most rapid decrease near a total dose of about $5\times10^{16}$ cm$^{-2}$. A $T_{cf}$ of 6.6 K is measured after exposing the sample to a total dose of $1.3\times10^{17}$ cm$^{-2}$. Again, $\Delta\rho_{300\text{-}41K}$ stays relatively constant, suggesting that the grains remain well connected. The transition width ($\Delta T_c = T_{co} - T_{cf}$) reaches a maximum when $T_{cf} \sim 15$ K at a dose of $7\times10^{16}$ cm$^{-2}$, where the $T_c$ decreases most rapidly with increasing dosage. The increase in the transition width is believed to be due to inhomogenities on the order of the coherence length[15] (~50 Å). Similar observations have been reported[10,11] for damage measurements on Nb$_3$Ge, Nb$_3$Sn, V$_3$Si, and V$_3$Ge.

Our data shed light on the two-band effects in the electrical transport properties. Mazin *et al.*[3] predict that samples in both the clean and dirty limit have similar temperature dependencies of $\rho$, and $\Delta\rho_{300\text{-}41K}$ is significantly larger in the dirty limit. They conclude that the increased resistivity in the dirty limit arises mostly as a result of an increase in the intraband π-band defect scattering, while the interband defect scattering and intraband defect scattering in the σ-band are small. Further, they explain their predicted small change in $T_c$ between the clean and the dirty limits by the exceptionally small interband scattering. Liu *et al.*[7] and Braccini *et al.*[8] have predicted that the $T_c$ of MgB$_2$ in the dirty limit would be about 20 K.

The $T_{cf}$ vs. $\rho(41K)$ resulting from different ion dosage levels are shown in Fig. 3. This $T_c$ vs. resistivity plot is linear for both film A and film B, similar to that of many A15 superconductors like Nb$_3$Ge and V$_3$Si.[11] The intercept on the resistivity axis represents the maximum intra-grain resistivity of an MgB$_2$ sample before superconductivity is destroyed, which is ~117 μΩcm for Film A and 72 μΩ.cm for Film B. Resistivity values in Film B might be less reliable as some cracking of the film near the voltage arms occurred after the third damage step. Before damage, the resistivity of the film is presumed to be determined by the π-band scattering as the slope d$\rho$/dT at 300K is 0.065 μΩcm/K, very close to the value of 0.06 μΩcm/K predicted for π-band conduction.[5] As damage proceeds, $\rho(41K)$ in film A increases by a factor of 55 when $T_c$ reaches 6.6K, but d$\rho$/dT(300K) changes only slightly. It is not possible to say with certainty whether the measured resistivity continues to represent only scattering in the π-band, or whether the σ-band and interband scattering contribute as well. The linear reduction of $T_c$ with resistivity, and lack of any saturation or plateau in $T_c$ near 20-25 K, suggests that other factors might dominate in reducing $T_c$. A likely explanation used, for example, to explain the $T_c$ of amorphous transition metals and damaged A15 superconductors, is that disorder smears and decreases the peak in the electron density of states at the Fermi level.[16] In the case of film B, the intercept of 72 μΩ.cm is similar to that in C-doped $MgB_2$,[17] where the $T_c$ was suppressed to 2.5 K at a resistivity of about 50 μΩ.cm.

We have performed similar experiments on a film grown by MBE[18] at ASU. The initial resistivity is much higher (~90 μΩ.cm) but the dependence of $T_c$ on the intragrain resistivity (derived from the model of reference 4) appears to be similar. These results, and measurements of the increase in $H_{c2}$ as $MgB_2$ films are damaged, will be published later.

To summarize, we have irradiated $MgB_2$ bridges using a 2-MeV $^4He^{++}$ ion beam. The results indicate a progressive damage of the grains themselves, leading to a reduction in $T_c$, while the connectivity between the grains remains unaffected. A plot of $T_c$ vs. $\rho$(41K) was found to be linear. No evidence is seen for saturation in $T_c$ near 20 K, predicted to be the dirty limit of interband scattering, suggesting that other factors determine $T_c$ in the damaged films. For a particular $T_c$, our study gives a measure of the intragrain resistivity in an $MgB_2$ sample. Thus $T_c$ is determined solely by the intragrain scattering in the sample, but as damage proceeds, we are not able to ascribe the resistivity specifically to π-band, σ-band, or interband scattering.

The work at ASU was supported by ONR under grant N00014-02-1-0002. We acknowledge use of facilities in the Center for Solid State Science at ASU. The work at Penn State was supported in part by ONR under grant N00014-00-1-0294 (Xi) and N0014-01-1-0006 (Redwing), by NSF under grant DMR- 0306746 (Xi, Redwing), and DMR-9876266 and DMR-9972973 (Li).

List of Figures:

FIG. 1 ρ-T curves at various damage levels for film A.

FIG. 2 Variation of $T_c$, $\rho(41K)$ and $\Delta\rho_{300\text{-}41K}$ with $^4He^{++}$ ion dosage for film A.

FIG. 3 $T_{cf}$ vs. residual resistivity plot, obtained from respective values observed at different damage levels for films A and B.

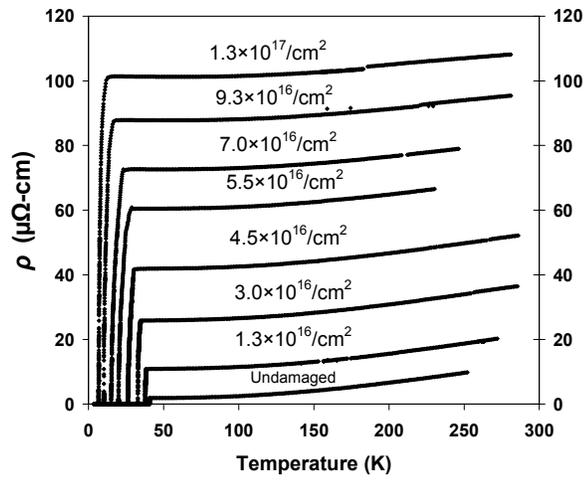

**FIG. 1- Gandikota et al.**

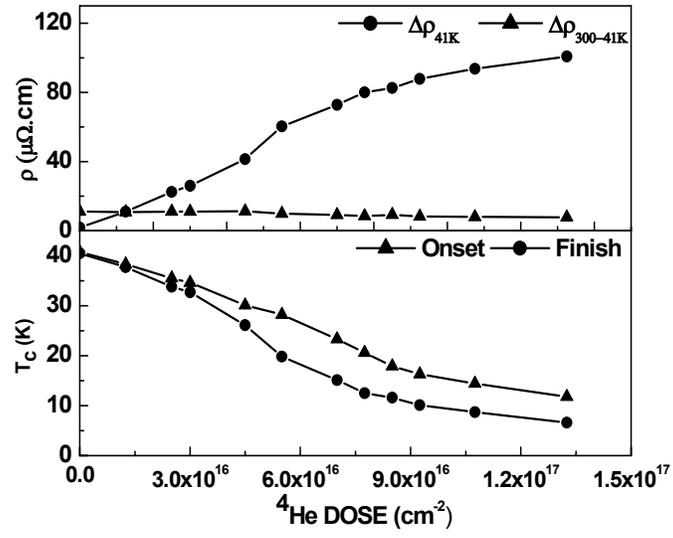

**FIG. 2- Gandikota et al.**

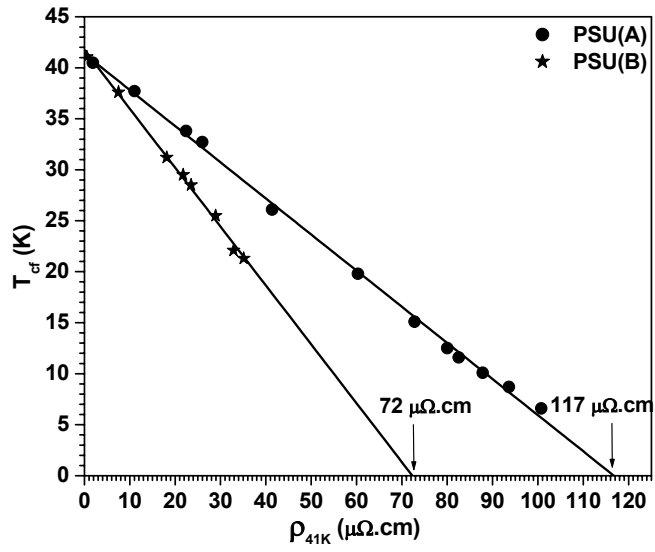

**FIG. 3- Gandikota et al.**